\documentclass{emulateapj}            
\newcommand{\ciii}{\ion{C}{3}}
\newcommand{\cii}{\ion{C}{2}}
\newcommand{\pv}{\ion{P}{5}}
\newcommand{\hii}{\ion{H}{2}}

\newcommand{\oiii}{\ion{O}{3}}
\newcommand{\nii}{\ion{N}{2}}
\newcommand{\zsun}{Z$_{\odot}$}
\newcommand{\msun}{M$_{\odot}$}
\newcommand{\ergs}{ergs s$^{-1}$ cm$^{-2}$ \AA$^{-1}$}
\newcommand{\ovi}{O~{\sc{vi}}}
\newcommand{\siv}{S~{\sc{iv}}}

\slugcomment{Submitted to Astronomical Journal}
\shorttitle{Massive Stars in Giant {\hii} Regions of M\,33 and M\,101}
\shortauthors{Anne Pellerin}

\begin{document}

\title{ {Massive Stellar Content of Giant {\hii} Regions in
         M\,33 and M\,101}\footnote{Based on observations made 
         with the NASA-CNES-CSA Far Ultraviolet Spectroscopic 
         Explorer. FUSE is operated for NASA by the Johns Hopkins 
         University under NASA contract NAS5-32985.} }    

\author{Anne Pellerin}
\affil{Space Telescope Science Institute,
        3700 San Martin Drive,
        Baltimore, MD, 21218, USA.}
\email{pellerin@stsci.edu}

\begin{abstract}

Far-ultraviolet (900-1200\AA) spectral synthesis of nine giant extragalactic 
{\hii} regions in M\,33 and M\,101 is performed to study 
their massive stellar content. Several parameters are quantified, 
predicted, and compared to the literature: age, stellar mass, IMF slope, 
number of O-type and Wolf-Rayet stars, H$\alpha$ and 5500\AA\ continuum 
fluxes. The results of this particular technique are consistent with other 
methods and observations. This work shows that a total stellar mass 
of a few {10$^3$\,{\msun}} is needed to populate the IMF bins well 
enough at high masses to obtain accurate results from the spectral 
synthesis technique in the far-ultraviolet. 
A flat IMF slope seems to characterize better the stellar line 
profiles of these objects, which is likely the first sign of a
small number statistic effect on the IMF. 
Finally, the {\hii} region NGC\,5461 is identified as a good 
candidate for hosting a second generation of stars, not yet seen at 
far-ultraviolet wavelengths. 

\end{abstract}


\section{Introduction}

Giant extragalactic {\hii} regions (GEHR) are important sites of star 
formation. They are small scale examples of extreme sites of star 
formation such as local and distant starburst galaxies. Like 
starburst regions, they contain several distinct star clusters 
\citep[e.g.][]{meu95,hun96} that can interact with each other to
potentially enhance or slow down the star formation processes
\citep[see review by][and references therein]{tan05}. They produce
the most massive stellar types known (O, B, and Wolf-Rayet) that have the 
potential to transform the morphological and chemical aspects of 
galaxies through their feedback 
\citep[e.g.][among others]{heck90,mar02,tre04,cal04}.
Most GEHR are recent and quasi-instantaneous events of star-formation
\citep{mas91,mas99,sch99,sta99}, as seems to be the case for a 
starburst, in the sense that most of their massive stars seems to
form within less than 2-3\,Myr \citep{pel04}.

Evolutionary synthesis is a powerful tool to study stellar populations in
various environments \citep[e.g.][]{wor94,lei99,bruz03,rob03}. The main goal 
of evolutionary synthesis is to deduce the global properties of spatially 
unresolved stellar populations such as their averaged 
age, mass, and metallicity. The development of evolutionary synthesis 
codes in the past decade has considerably improved our knowledge of 
galaxies \citep[e.g.][among many others]{gonz99,lei01,chan03}. 
With the recent (and coming) generation of large 
telescopes such as Keck, Gemini, JWST, and ALMA, this technique will be 
very useful for our understanding of very distant galaxies
and of their evolution through cosmic time.

Nearby GEHR are excellent candidates to test the 
accuracy of the evolutionary synthesis technique. GEHR like those found 
in M\,33 are close enough to resolve individual stars and to compare their 
detailed stellar content with what is deduced from synthesis of
integrated spectra. In this work, a detailed study of the massive stellar 
content of several GEHR observed in the far-ultraviolet (900-1200\AA; FUV)
is presented. The study is based on the
spectral synthesis code {\tt LavalSB} and its recent empirical spectral
library in the FUV range \citep{rob03}. 
The synthesis of GEHR observed in M\,33 and M\,101 will be compared, when
possible, to previous works detailing their resolved stellar content.

The following section presents a summary of the data processing. 
Section~\ref{lavalsb} describes the evolutionary synthesis code 
{\tt LavalSB} used in this work. The synthesis results for each GEHR 
are detailed in \S\ref{syn}, and compared with previous works at various 
wavelengths. A discussion of specific results is presented in section~5
and the main results are summarized in \S6.

\section{FUSE Data and Reduction}
\label{data}

FUV spectrograms of nine GEHR were obtained by the {\it {Far Ultraviolet
Spectroscopic Explorer}} (FUSE) telescope \citep{moos00} for various 
projects. Most data were obtained through the largest aperture
(LWRS; 30$^{\prime\prime}$$\times$30$^{\prime\prime}$) while some
spectrograms were obtained using a smaller aperture (MDRS; 
4$^{\prime\prime}$$\times$20$^{\prime\prime}$). Aperture locations
are displayed in Figure~1.
A general description of the FUSE data is reported in Table~1.
Data were gathered from the MAST\footnote{Multimission 
Archive at Space Telescope Science Institute; http://archive.stsci.edu/\,.} 
public archives.  The data were processed with the {\tt calfuse} pipeline 
v2.4.2. This version corrects for Doppler shift induced by the 
heliocentric motion of Earth, event bursts, the walk problem, 
grating thermal shifts, bad pixels, background noise, distortions, and 
astigmatism. More information relative to {\tt calfuse} is available
electronically\footnote{http://fuse.pha.jhu.edu/analysis/calfuse.html}.
The output from {\tt calfuse} comprises eight segment spectrograms for each
exposure that correspond to the eight optical paths of the instrument. 
Each segment covers a different wavelength range, with some of them 
overlapping \citep[see fig.~2 of][]{sah00}. First, for each segment, 
each exposure was combined with a statistical weight based on 
exposure time. Then the segments that cover the same wavelength 
regions (roughly 900-1000\AA, 1000-1100\AA, and 1100-1200\AA) were
averaged with weights based on their signal-to-noise ratios. Finally, 
the spectrograms of each wavelength range were simply coadded to 
obtain one spectrogram covering the entire 905-1187\AA\ range. The 
spectrograms were then smoothed by a factor of 20 using the 
IRAF\footnote{Image Reduction and Analysis Facility, supported by NOAO 
and operated by AURA under cooperative agreement with the 
NSF; http://iraf.noao.edu/\,.} {\it {boxcar}} task, corresponding to a 
resolution of about 0.13\AA. This last step increases the signal-to-noise 
ratio without affecting the stellar line profiles. The spectrograms were
corrected for redshift. Reddening correction will be
discussed in section~\ref{syn}.

\section{Stellar population modeling in the FUV}
\label{lavalsb}

A first work of spectral synthesis below 1200\AA\ has been made
by \citet{gonz97} for the {\ovi}$+$Ly$\beta$$+${\cii} feature. The 
stellar library was based on Copernicus and the Hopkins Ultraviolet 
Telescope (HUT) with a spectral resolution of 0.2\AA. Their work 
clearly showed that the line profile was sensitive to the age of 
a stellar population.
A new FUV spectral library, based on FUSE data, has recently been added 
to the spectral synthesis code {\tt LavalSB} \citep{rob03}. This code has
been proven to be very powerful for young stellar populations 
\citep{pel04} and will be used in the present work to deduce the global 
properties of massive stars in GEHR from their integrated FUV light. 
{\tt LavalSB} is a parallel version of {\tt Starburst99} \citep{lei99}. 
It uses the evolutionary tracks of the Geneva 
group (Schaller et al. 1992; Schaerer et al. 1993a, 1993b; 
Charbonnel et al. 1993; Meynet et al. 1994).
The stellar population follows a mass distribution based on a chosen 
stellar initial mass function (IMF) and mode of star formation
(instantaneous or continuous). Individual stellar parameters are used
to assign the corresponding normalized empirical spectrogram from the FUV 
library based on relations from \citet{schm82}. The normalized library 
spectrograms are flux calibrated using stellar atmosphere models of 
\citet{kur92} for normal stars, and of \citet{sch92} for stars with 
extended envelopes. 
The \citet{kur92} spectra have been fitted using a Legendre function to 
remove their low resolution spectral features in order to avoid
any confusion with empirical stellar lines from the spectral library.
The FUSE stellar library covers from 1003.1 to 1182.678\AA\ with a 
dispersion of 0.127\AA. The library metallicities corresponds to the
evolutionary tracks of {\tt LavalSB}, e.g. {\zsun} for 
Galactic stars \citep[{12$+$log[O/H]$=$8.7};][]{all01}, 0.4\,{\zsun} 
for LMC stars \citep[{12$+$log[O/H]$=$8.3};][]{rus92}, and 0.1\,{\zsun} 
for SMC stars \citep[{12$+$log[O/H]$=$8.0};][]{rus92}. 

The most useful stellar indicators in the FUV are the {\ciii} blend 
multiplet centered at 1175.6\AA, and the {\pv} doublet at 1118.0 and 
1128.0\AA. The profiles of these lines show strong variations with age 
and metallicity of the population, depending on what spectral types 
dominate in flux. Significant, but more subtle, changes 
also appear with different IMF parameters. At shorter wavelengths, the 
{\ovi~$\lambda\lambda$1031.9, 1037.6} and the {\siv~$\lambda\lambda$1062.7, 
1073.0, 1073.5} line profiles show variations with age and metallicity, and
possibly with IMF. However, the empirical stellar library used in 
{\tt LavalSB} contain stars for which these diagnostic lines are 
contaminated by interstellar features from Galactic H$_2$ and other 
atomic transitions. Consequently, stellar lines of {\ovi} and {\siv}
will not be used in the present work since {\ciii} and {\pv} lines alone
will provide more accurate results. An extensive identification 
of stellar and interstellar lines contained within the FUSE range can be
found in \citet{rob03} and \citet{pel02}.

To establish the characteristics of an integrated stellar population in 
the FUV, the FUSE spectrogram is first normalized and the stellar indicators 
of {\ciii~$\lambda$1175.6} and {\pv~$\lambda\lambda$1118.0, 1128.0} are
compared to the models. The best-fit model is chosen both by eye and by
performing a $\chi^2$ fit. This first step provides information on the age, 
the metallicity, and the IMF parameters of the population. A standard IMF
is defined here as having a slope $\alpha$=2.35 a mass range from 1 to 
100\,{\msun}. Once the age and metallicity of the stellar population are 
estimated
from the normalized FUV spectrogram, the extinction is then evaluated by 
comparing the observed continuum slope of the flux calibrated data to
the one of the best-fit model. The theoretical law from \citet{witt00} 
for a clumpy dust shell distribution with an optical depth of 1.5 in the 
V band is used to derive the internal extinction E(B-V)$_i$. The Galactic 
extinction is corrected using the law of \citet{sea79}. Finally, the 
stellar mass involved in the system is estimated from the unreddened 
flux level.

Uncertainties related to the line profile fitting are determined by
comparing the different sets of models at a given metallicity. Since
{\tt LavalSB} covers only specific values 
(0.1\,{\zsun}, 0.4\,{\zsun}, {\zsun}, and 2\,{\zsun}), 
it is not possible at this point to evaluate the full age range 
that could fit the data. The jumps in metallicity are quite large so
the synthetic spectra from the next metallicity value do not always 
reproduce the observed line profiles and cannot give
clues on the age range. Consequently, the age uncertainties given 
in the present work are underestimated and do not 
take into account the possibility that the data can be fitted using a 
slightly different age at a slightly different metallicity.

For a given model, the primary source of error in the estimation of 
stellar masses and predicted fluxes is usually the FUV flux uncertainty 
from FUSE, which is usually around 10\%. However, in some cases, the 
age uncertainty gives a larger error bar than
the FUSE uncertainty. In every case, the largest uncertainty is
given. Also, the IMF slope used to calculate the total stellar mass 
affects the uncertainty on masses and predicted fluxes. However, 
these uncertainties are not explicitly
included for the best fit model uncertainties. Where possible, 
parameters of other good-fit models are given to better evaluate 
the full uncertainties.

\section{Massive Stellar Content of GEHR: the FUV Point of View}
\label{syn}

\subsection{NGC\,604}
\label{n604}

NGC\,604 is a well-known GEHR within the Local Group galaxy M\,33. 
Several studies found and confirmed the presence of very massive
O, B, and Wolf-Rayet (WR) stars \citep{vil88,dri93,hun96,gonz00,bru03,
maiz04}. At least four distinct starclusters have been identified in this object
\citep{hun96}. 

The FUSE spectrogram of NGC\,604 obtained through the LWRS aperture is 
shown in Figure~2a. This aperture corresponds to a physical size of 
123$\times$123\,pc$^2$ \citep[1$^{\prime\prime}$=4.1\,pc at 
840\,kpc; see also Fig.1 of][]{leb05}. The aperture includes the 
Cluster~A from \citet{hun96}, but not the entire {\hii} region. The 
spectrogram has a very good signal-to-noise ratio (S/N) of 20 
between 1155 and 1165\AA\ that allows to perform a good synthesis 
with details on the IMF slope. The {\ciii} line profile shows a large 
absorption feature in its blue wing, indicating the presence of evolved 
late-type O~stars. The {\pv} doublet 
also displays P\,Cygni line profiles typical of massive stars with
strong winds. The line depths of {\ciii} and {\pv} suggest a sub-solar 
metallicity for the stars. Continuous burst models have to be excluded 
since they produce stellar lines with too faint P\,Cygni profiles. To
obtain a good fit, especially for the {\ciii} line profile, a flatter IMF
with a slope $\alpha$=1.5-2.2 is better, while a standard IMF with 
$\alpha$=2.35 could also fit. The best fits are obtained for models
having $\alpha$=1.5 and an age of 3.9$\pm$0.1\,Myr for 0.1\,{\zsun} and 
3.3$\pm$0.1\,Myr for a 0.4\,{\zsun} metallicity. If $\alpha$(IMF)=2.35, 
then the best-fit ages are a little lower with 3.5$\pm$0.3\,Myr for 
0.1\,{\zsun} models and 3.0\,Myr at 0.4\,{\zsun}. 
The  solution is not unique since there is a degeneracy in the 
line depth for the models at sub-solar metallicities when 
the P\,Cygni profiles are well developed.

\citet{gonz00} performed a detailed study of NGC\,604 using IUE spectrograms
(9.5$^{\prime\prime}\times$22$^{\prime\prime}$ aperture), optical
ground-based data, and H$\alpha$ images from the HST, to fully describe 
this GEHR. From the H Balmer and {\ion{He}{1}} absorption lines, they 
deduced an age between 3 and 4\,Myr for the stellar population with a 
standard IMF or flatter. A continuous burst cannot fit their 
emission line ratios. Their IUE spectrograms revealed a population of 
3-5\,Myr (better fit at 3\,Myr) with an IMF slope flatter than 3.3. 
\citet{vil88} studied in detail the chemical abundances in M\,33 from 
nebular lines. They measured an oxygen abundance 12$+$log[O/H]=8.51 for
NGC\,604. All these results are fully consistent with the FUV line profile 
synthesis. The best-fit model parameters are reported in Table~2, 
together with the other good-fit models. Note that hereafter, 
calculations using the models at 0.4\,{\zsun} are favored based on 
the metallicity from \citet{vil88}. 

Adopting an instantaneous burst model of 3.3\,Myr at 0.4\,{\zsun} with 
an IMF slope $\alpha$=1.5, the observed FUV continuum slope suggests no 
significant internal extinction E(B-V)$_i$. No internal extinction is needed if a 
Galactic correction of 0.02 is applied, and E(B-V)$_i$=0.03 is calculated 
if no Galactic extinction is applied. Using an IMF truncated between 1 
and 100\,{\msun}, the FUV flux level leads to a stellar mass of 
(7$\pm$2)$\times$10$^3$\,{\msun} within the LWRS aperture. 
Using an IMF slope of 2.35, the calculated stellar mass is 
rather (1.4$\pm$0.3)$\times$10$^4$\,{\msun}. \citet{gonz00} obtained a 
E(B$-$V)$_i$ of 0.1 based on their IUE spectrograms.
They also estimate a stellar mass of 0.1-2$\times$10$^5$\,{\msun}. 
\citet{hun96} found, based on optical HST images, an extinction value 
of 0.08 for Cluster~A contained within the LWRS aperture.
Their extinction and mass values are slightly higher than those from the
FUSE data.

From the stellar population described above with $\alpha$=1.5, several 
physical parameters can be deduced and compared (see Table~3). First, 
such a population would theoretically lead to an unreddened H$\alpha$ 
flux of (2$\pm$1)$\times$10$^{-11}$\,{\ergs}, and a continuum level at 
5500\AA\ around (3$\pm$1)$\times$10$^{-12}$\,{\ergs}.
Changing the IMF does not change these numbers significantly.
H$\alpha$ fluxes of 4.0 and 3.3 $\times$10$^{-11}$\,{\ergs} have 
been measured from HST and ground-based images by \citet{gonz00} and 
\citet{bos02}, respectively. Those values are slightly above the FUV 
predicted values. The differences in H$\alpha$ fluxes are consistent 
with the differences in stellar masses. 

According to
the H$\alpha$+UV images from HST \citep[see Fig. 2 of][]{gonz00}, several
massive stars from Cluster~A are co-spatial with the nebular emission.
These stars are good candidates to higher extinction and it is likely
that their contribution to the FUV flux is significantly lower than at
longer wavelengths (even at $\sim$1500\AA) and then partly explain 
the differences observed in extinction values at various 
wavelength ranges. Furthermore, Fig.~2 of \citet{hun96} shows that
Cluster~B and C contribute significantly to the nebular emission of 
NGC\,604, but they are not taken into account in the total stellar 
mass derived from FUV since they are not included within the FUSE 
aperture. Also, a detailed study from 
\citet{maiz04} revealed an extremely complex gas/dust geometry for which
around 27\% of the ionizing photons might be missing in NGC\,604 due 
to attenuation. In addition to the aperture effect, this obviously 
contributes to create a discrepancy between the predicted and observed
values in the stellar mass and other fluxes parameters.

The FUV synthesis of a 3.3\,Myr population with $\alpha$=1.5 at 0.4\,{\zsun}
predicts that about 9 WR stars (3 WN and 6 WC) should be 
present in NGC\,604. \citet{dri93} obtained ground-based and HST-WF/PC1 
images and identified 12 WR or Of candidates, slightly more than 
the {\tt LavalSB} predictions. More recently Drissen et al. 
(2005, in preparation) confirms that there are at least 6 WN and 2 WC
stars among them. This WC-to-WN number ratio is not consistent with 
{\tt LavalSB} (or {\tt Starburst99} neither). 
To obtain WC/WN$\sim$1/3, both models propose an age around 4.5-4.7\,Myr 
for the population. However, {\tt LavalSB} do not 
include the effect of rotation in evolutionary tracks.
By including rotation in the models,
the result will be to extend the duration of the 
WR phase and to increase considerably the number of WN stars, which will
fit better the observations (G. A. V\'azquez 2005, private communication).

Also, for the population synthesized above for NGC\,604, LavalSB predicts that 
90$^{+30}_{-10}$ O-type stars (of all spectral types still present 
at this age). \citet{hun96} estimate from 
HST/WFPC2 images that about 190 stars brighter than O9.5\,V are present 
in NGC\,604, which is higher than the FUV estimation. However,
the number of \citet{hun96} may include some B supergiants. If 
we use an IMF slope of 2.35, the model then predicts roughly the same number 
of O-type stars but no WR stars (or very few) at 3.0\,Myr, which 
is in disagreement with the observations of \citet{dri93}. The comparison
between the predicted and observed number of WR stars favors the case of
an IMF slope flatter than 2.35.

The FUSE spectrogram of the inner part of NGC\,604 obtained through the 
MDRS aperture is shown is Figure~2b (S/N$\sim$14). This smaller 
aperture corresponds to a physical size of 16$\times$82\,pc$^2$. The 
stellar line profiles are similar to those obtained with the LWRS 
aperture, but not exactly the same. The {\ciii} and {\pv} 
line profiles cannot be reproduced as well as for the LWRS data, 
especially in their blue wings. The models closer to the observed line
profiles are those of 3.9-4.1\,Myr at 0.1\,{\zsun} and 3.3-3.4\,Myr using 
0.4\,{\zsun} models. Interestingly, the MDRS spectrogram of NGC\,604 
corresponds better to a combination of a synthesized population and 
the spectrogram of a O8\,I LMC star. The blue wings in {\pv} and {\ciii} 
profiles are fitted by the single star spectrogram,
while the photospheric portion cannot be fitted by the star, but by 
a modeled population. This strongly suggests that
the number of massive stars within the aperture is low enough to be
subject to statistical biases on the stellar IMF, and is not well
represented anymore by an analytical IMF. Assuming a stellar 
population of 3.3\,Myr at 0.4\,{\zsun} as found previously, the 
continuum slope for the MDRS spectrogram gives E(B-V)$_i$ 
0.03$\pm$0.02 if no Galactic extinction is considered. 
The flux level indicates a stellar mass of about 
1$\times$10$^{3}$\,{\msun} through the MDRS
aperture, clearly indicating that the MDRS aperture does not include the
whole GEHR.

\subsection{NGC\,595}
\label{n595}

As NGC\,604, NGC\,595 contains multiple star clusters with OB stars 
\citep[e.g.][]{dri93,mas99,maiz01}. The FUSE spectrogram of NGC\,595 is 
presented in Figure~2c with S/N$\sim$13. Particularly strong P\,Cygni 
profiles are observed in {\ciii} and {\pv}. As for a single evolved O~star, 
the {\ciii} profile of NGG\,595 does not show a blend of 
photospheric$+$wind features as in an integrated population, but a 
single well-developed P\,Cygni profile. In fact, it appears that a synthesized stellar 
population is unable to reproduce the FUV line profiles. The FUSE 
spectrograms have then been compared to those of single O stars from 
the FUV stellar library of {\tt LavalSB} and it reveals that an O7\,I 
LMC star is the closest match to the spectrogram of NGC\,595 (see 
superimposed thick line spectrogram in Figure~2c). 
It is obvious here that there are not enough hot stars in NGC\,595 to 
fit an analytical IMF as used in current spectral synthesis. Only a few 
stars with strong winds seem to dominate the line profiles. 

According to {\tt LavalSB}, O7\,I stars appear between 2.5 and 
4.0\,Myr after an instantaneous burst. At 2.5\,Myr, stars slightly 
brighter than O7\,I will probably dominate the FUV flux.
Consequently, the O7\,I stars in NGC\,595 would be consistent with an age of
3.5$\pm$0.5\,Myr with a metallicity close to the LMC (0.4\,{\zsun}). This age 
is consistent with the works of \citet{mal96} and \citet{mas99}. Assuming a 
standard IMF, it is still possible to roughly estimate parameters related to the
FUV slope and flux level. Adopting a Galactic extinction of 0.04 
(NED\footnote{The NASA Extragalactic Database (NED) is operated by the 
Jet Propulsion Laboratory, California Institute of Technology, under 
contract with the National Aeronautics and Space Administration; 
http://nedwww.ipac.caltech.edu/\,.}), a very low internal extinction of 
E(B-V)$_i$=0.02$\pm$0.02 is found. The stellar 
mass of NGC\,595 is then estimated to be about 1$\times$10$^3$\,{\msun}
with very large uncertainties. 
Previous works in the visible range suggested a higher extinction value of 
0.3 \citep{mal96,mas99,maiz01} for this GEHR. \citet{mas99} and \citet{mal96} 
also estimated a stellar mass of 5-6$\times$10$^3$\,{\msun}, which is also
significantly higher than the FUV result, but of the same order of magnitude. 

Based on {\tt LavalSB}, the age and mass of NGC\,595 suggest that about 10
O~stars and 1 or 2~WR should be present in NGC\,595. However, HST 
imaging reveals larger numbers of these stars. \citet{dri93} identified 
11~WR/Of candidates and \citet{mal96} estimated the number 
of O~stars to be $\sim$90. \citet{dri93} estimate that there are 2.5 times fewer stars 
between 15 and 60\,{\msun} in NGC\,595 than NGC\,604, implying that NGC\,595 
must be about 2.5 times less massive than NGC\,604. FUV synthesis gives a 
factor of 5 between the stellar masses of the two GEHR. Recently, optical spectra 
from Drissen et al. (2005, in preparation) confirmed the presence of several WR 
candidates within NGC\,595 and classified them. Based on the
HST/WFPC2-F170W archival image, the WR stars produce about 30\% of 
the UV luminosity. Obviously, the observed number of WR stars in this object is 
incoherent with the FUV synthesis point of view.

In an attempt to reproduce the observed FUV spectrogram, simple combinations of 
individual hot stars are tested. The combinations are comprised of individual late 
O-type stars (or synthetic models) and WR stars for which $\sim$ 30\% of the total FUV 
flux comes from 1~WN6/7 star and 4~WN7/8 stars, as classified by Drissen et al. 
(2005, in preparation). However, the resulting fits are poor, with the stellar combinations 
always giving wind profiles too strong in emission and having too narrow blue absorption. 
However, the FUSE atlas of WR stars from \citet{wil04} revealed spectra of 
WR stars in general with spectral line profiles that are changing considerably from one type to
another. A closer look at this atlas shows that HDE\,269927, a WN9 type star from the
Galaxy, display line profiles of {\ciii} and {\pv} similar to stellar lines of NGC\,595. Replacing
the WN7/8 spectra used in the previous combinations by the spectra of HDE\,269927
gives surprisingly good results. In fact, the combination of spectrograms from a O7\,I star (70\%
of the flux) as well as 1 WN6 and 4 WN9 stars (30\% of the flux) reproduces well the FUSE data 
for NGC\,595. This implies two things. First, it appears that the FUV spectra of WR stars show
line profiles that change significantly from one spectral type to another, and that probably vary
with metallicity as well. Consequently, the few WR spectrograms currently used in the {\tt LavalSB}
spectral library are probably not very representative of their spectral types. Fortunately, these
stars do not usually contribute significantly to an integrated stellar population and then do not 
really affect the synthetic spectra. Second, it seems obvious that the FUV spectra of NGC\,595 
is dominated by evolved late-type O and WN-late stars. However, one fundamental question 
remains: how did NGC\,595 come to produce a stellar population enhanced in WR stars?

The FUV synthesis of NGC\,595 implies that 
F(H$\alpha$)=(1.3$\pm$0.2)$\times$10$^{-12}$\,{\ergs}. Various values 
are found in the literature. \citet{bos02} obtained 
1.1$\times$10$^{-11}$\,{\ergs}, and \citet{ken79} measured 
8.8$\times$10$^{-12}$\,{\ergs}. It is obvious that the FUV synthesis 
is not accurate in this case, and possibly also that it does not include the entire 
GEHR.

The FUSE spectrogram of NGC\,595 clearly reveals that a stellar population 
with a stellar mass of a few 10$^3$\,{\msun} is too small to apply the 
spectral synthesis technique, at least below 1200\AA. Obviously, 
statistical fluctuations related to a small number of massive stars 
are not well represented by an analytical IMF. A more detailed discussion
on this subject will be given in \S\ref{mass}

\subsection{NGC\,592}
\label{n592}

Because of its fainter H$\alpha$ luminosity, NGC\,592 is a much less 
studied GEHR, but not less interesting. The observed FUV spectrogram is 
shown in Figure~2d, with a rather low S/N of 6. The FUSE aperture 
contains the entire GEHR \citep{bos02,keel04}. Despite noisy stellar 
lines, their profiles clearly display extended blue absorption wings 
from evolved O stars. Comparing both {\pv~$\lambda$1128.0} and 
{\ciii~$\lambda$1175.6} lines to the models, it is possible to reproduce 
their profiles with a 4.0$\pm$0.5\,Myr stellar population at {\zsun} 
metallicity. Models at 0.4\,{\zsun} produce too weak P\,Cygni effects in 
{\ciii}. The spectrogram is too noisy to discriminate between various IMF
slopes. From H$\alpha$ and H$\beta$ narrow-band
images, \citet{bos02} estimated the age of NGC\,592 to be more than 
4.5\,Myr, which is not really compatible with FUV line profiles displaying
relatively strong P\,Cygni features. In term of metallicity, \citet{keel04} 
interpolated a value of 0.5\,{\zsun} in [O/H], and Drissen et al. (2005, in
preparation) estimated that 12$+$log[O/H]$\sim$8.4 (i.e. 0.5\,\zsun) 
from [\oiii]/H$\beta$ and [\nii]/H$\alpha$ 
line ratios. These values are consistent with the FUV synthesis 
considering that {\zsun} models can cover relatively well a metallicity 
range from 0.4-0.5 to $\sim$1.2\,{\zsun} \citep{pel04}.

Using a model of 4.0\,Myr at {\zsun} and a standard IMF, and assuming a
Galactic extinction of 0.042 (NED), an E(B-V)$_i$ of 0.07$\pm$0.02 is 
deduced from the FUV continuum slope. Once the data are corrected 
for extinction, the
stellar mass deduced is (1.1$\pm$0.3)$\times$10$^4$\,{\msun}. This mass is
similar to that estimated for NGC\,604, which is consistent with the
fact that the stellar line profiles can be reproduced with a synthesis 
technique and an analytical IMF, contrary to NGC\,595. The FUV flux 
level implies a unreddened H$\alpha$ flux of 
(2.7$\pm$0.5)$\times$10$^{-12}$\,{\ergs}, which is the exact 
value by \citet{bos02}. Other predicted parameters are 
reported in Table~3.

\subsection{NGC\,588}
\label{n588}

The FUSE spectrogram of NGC\,588 is presented in Figure~2e, with a good S/N
of 12. The FUSE aperture includes the entire {\hii} region 
\citep{bos02,keel04}. Models at {\zsun} produce stellar lines definitely 
too deep compared to the observations. With models at 0.4\,{\zsun} 
metallicity, a good fit can be obtained for a 3.5$\pm$0.5\,Myr population 
with $\alpha$(IMF)$\leq$2.35. A flatter IMF tends to give better results, 
but it is hard to really distinguish between various IMF slopes because of 
the relatively low S/N. Good fits can also be obtained with 0.1\,{\zsun} 
models of 4.5$\pm$1.0\,Myr and still with $\alpha$(IMF)$\leq$2.35. In 
the literature, ages of 2.8, $>$4.5, and 4.2\,Myr are reported for 
NGC\,588 \citep[][respectively]{mas99,bos02,jam04}, in general 
agreement with FUV line profiles. \citet{vil88} derived a precise 
oxygen abundance of 12+log[O/H]=8.30 (i.e. 0.4\,{\zsun}), favoring
the models at 0.4\,{\zsun}. A flat IMF is also favored by \citet{mas99} 
and \citet{jam04} obtained $\alpha$(IMF)=2.37$\pm$0.16 from a star
counting method.

Based on the best-fit model at 0.4\,{\zsun}, a low internal extinction 
of at most 0.06$\pm$0.02 is measured, which leads to a stellar mass of 
(1.3$\pm$0.6)$\times$10$^3$\,{\msun}. The mass is higher,
(4$\pm$1)$\times$10$^3$\,{\msun}, if we consider $\alpha$=2.35. 
Depending on the extinction law used, E(B-V)$_i$ values between 0.11 
and 0.08 are measured \citep{mas99,jam04}. These same authors obtained 
stellar masses of 534 and 3000-5800\,{\msun}, respectively. The smallest 
value was deduced from IUE data (aperture of 
10$^{\prime\prime}\times$20$^{\prime\prime}$), and the largest mass 
is from full field imaging data, which explains the discrepancy.
FUV data are in relatively good agreement with imaging data, 
which suggests that most OB stars of NGC\,588 are within the FUSE 
aperture. With such a mass, the model predicts that 
F(H$\alpha$)=2.8$\times$10$^{-12}$\,{\ergs}, which is in good 
agreement with the value of 2-3$\times$10$^{-12}$\,{\ergs} measured by 
\citet{ken79} and \citet{bos02}. The best-fit model predicts 2 WR stars
in NGC\,588, which is the exact number found by \citet{jam04} in their
HST images with resolved stars.

\subsection{NGC\,588-NW}

A FUSE spectrogram has been obtained in the vicinity of NGC\,588 
(North-West). From the {\it {Digitized Sky Survey}} image (see Fig.~1), 
this region corresponds to a relatively compact and small 
cluster with a faint, extended nebular ring. It was first reported by 
\citet[][their object 281]{bou74} and also identified in the 
work of \citet{cou87}. The ring suggests that the cluster is more evolved
than those synthesized above. The FUSE spectrogram for this cluster is shown 
in Figure~2f (S/N$\sim$7). Diagnostic stellar lines do not display P\,Cygni 
profiles. Synthetic models do not reproduce well the line profiles. The 
best fit is obtained for a stellar population around 5-6\,Myr old at 
0.4\,{\zsun}, but the line profiles are not properly fitted. A possible 
alternative is a single star spectrum, as was the case for NGC\,595. Then, 
a Galactic O9.5\,III star also consistent with a population of 5-6\,Myr, 
gives a better match than the model but significant discrepancies still 
exist. This age is consistent with the presence of the faint extended 
ring seen around NGC\,588-NW in the visible range.

To push the synthesis further, a stellar population of 5.5\,Myr at 
0.4\,{\zsun} has been considered and an extinction value around 0 and a 
stellar mass of about 1$\times$10$^3$\,{\msun} have been roughly 
estimated for this cluster. This stellar mass is similar to the one 
obtained for NGC\,595. The relatively low mass of the cluster is 
a logical explanation for why the synthesis technique does not work 
well. Rough estimations of predicted observable parameters are 
reported in Table~3.

The study of NGC\,588-NW gives some other clues on the evolution of GEHR. 
First, the FUSE spectrogram of NGC\,588-NW reveals the presence of an 
important stellar population. However, because of 
its slightly greater age (5-6\,Myr instead of $\sim$3.5\,Myr for NGC\,595), 
the nebular emission is not as strong as for NGC\,595 and this region is 
consequently much less studied. It is likely that NGC\,588-NW is 
representative of what NGC\,595 may look like in $\sim$2-3\,Myr.
Second, the GEHR is still young and massive enough at this age not to 
have dissolved yet into the galaxy background. It would be interesting to search 
for slightly more evolved GEHR to better study their evolution, such 
as the dissipation timescale of clusters. This kind of cluster (i.e.
still very young but with significantly low nebular emission) may be at 
the origin of the diffuse UV light in starburst galaxies \citep{meu95}. 
NGC\,588-NW is consistent with clusters of less than 10$^3$\,{\msun} without 
O-type stars, as described by \citet{chan05} for the diffuse UV component in 
starbursts. A more extensive search for this kind of object in local galaxies 
could settle this issue.

\subsection{NGC\,5447}
\label{n5447}

NGC\,5447 is a GEHR in the spiral galaxy M\,101 (7.4\,Mpc) that displays 
several knots of star formation \citep{bos02}. The FUSE spectrogram has a 
S/N of 12 and is shown in Figure~3a. As shown in Fig.~1, the FUSE aperture 
does not include all knots. The spectrogram does not show strong 
wind profiles, suggesting that 
most O stars have already disappeared. Models at {\zsun} metallicity 
produce too deep stellar lines compared to the observations. Models at 
0.1\,{\zsun} cannot reproduce both {\pv} and {\ciii} features at the same 
age. The best-fit model is obtained at 4.5$\pm$0.5\,Myr with an IMF slope 
of 2.35 or flatter. This GEHR has not been extensively 
studied and no age has been proposed so far for this object. \citet{sco92} 
deduced an oxygen abundance of 8.3 in 12+log[O/H], compatible with the
line depths of {\pv} and {\ciii}.

The measured FUV slope for NGC\,5447 suggests that E(B-V)$_i$=0. From
photographic plates and the Balmer decrement, \citet{smi75} estimated an
extinction of 0.37, much larger than the FUV value. The FUV flux 
indicates a stellar mass of (1.2$\pm$0.2)$\times$10$^5$\,{\msun}.
From FUV synthesis, {\tt LavalSB} predicts that 
F(H$\alpha$)=(5.7$\pm$0.9)$\times$10$^{-13}$\,{\ergs}, 
and an EW(H$\alpha$)=1064\AA\ for 
NGC\,5447. Using photometric data \citet{bos02} measured an H$\alpha$ 
flux of 4.7$\times$10$^{-12}$\,{\ergs}, and \citet{ken79} obtained a 
value of 1.6$\times$10$^{-12}$\,{\ergs}. Since the GEHR is much more
extended than the FUSE aperture \citep[see Fig.~5 of][]{bos02}, 
the factor 5-10 discrepancies can easily be explained. However, the presence of 
a second generation of stars contributing to the nebular flux but not to 
the FUV flux cannot be excluded (see \S\ref{2egen}). For their knot~A only, 
\citet{bos02} obtained that F(H$\alpha$)= 7.5$\times$10$^{-13}$\,{\ergs},
suggesting that this knot must be the principal contributor to the FUV flux
measured with FUSE. \citet{tor89} measured a dereddened equivalent width of 
1096\AA\ through a 3.8$^{\prime\prime}\times$12.4$^{\prime\prime}$ slit, in 
very good agreement with the FUV predictions and the knot~A.

\subsection{NGC\,5461}
\label{n5461}

NGC\,5461 is a very large GEHR ($>$500\,pc in diameter) with multiple 
components in M\,101 \citep{bos02,keel04,chen05}. The FUSE aperture contains 
most of the H$\alpha$ emission and should include most of the massive
stellar content (see again Fig.~1). The FUSE spectrogram is shown in 
Figure~3b, with a S/N of about 7. The {\ciii} feature displays a wind 
profile, implying the presence of giant and supergiant {O-type} stars. 
Models at {\zsun} do not reproduce the stellar line depth. The models at 
0.1\,{\zsun} give a good fit for a 4.0$\pm$0.2\,Myr stellar population and 
an IMF slope flatter than 2.35. A good correspondence is also obtained with 
0.4\,{\zsun} models at 3.3$\pm$0.2\,Myr, still with $\alpha$$<$2.35. A 
multiwavelength study from \citet{ros94} suggests an age between 3.0 and 
4.5\,Myr, compatible with FUV line profiles. \citet{lur01} deduced an age 
between 2.5 and 3.5\,Myr based on EW(H$\beta$), also in general agreement 
with FUV line profiles. While the age determination method using 
EW(H$\beta$) is not a recommended diagnostic \citep{ter04}, it appears 
that it still gives good results at a such very young age.
More recently, \citet{chen05} identified about 12 candidate
stellar clusters within NGC\,5461 of which half of them are less than 
5\,Myr old. The other clusters are probably older and do not seem to
contribute much to the FUV flux. Abundances ranging from
8.4 to 8.6 in 12+log[O/H] are found in the literature 
\citep{tor89,sco92,ros94,lur01}. Their observations favor the FUV 
synthesis models at 0.4\,{\zsun}.

Comparing with the modeled population of 3.3\,Myr at 0.4\,{\zsun} and
$\alpha$=1.5, the FUV continuum slope needs no extinction correction. 
The stellar mass is then (1.5$\pm$0.4)$\times$10$^{4}$\,{\msun}. Using 
a standard IMF slope of 2.35, the calculated stellar mass is then
(5$\pm$1)$\times$10$^4$\,{\msun}. According to \citet{ros94}, the 
extinction from the Balmer decrement is 0.23, and using an extinction 
law especially designed for M\,101, they find a stellar mass of
1$\times$10$^{5}$\,{\msun}. 
According to {\tt LavalSB}, the FUV stellar population should
produce an H$\alpha$ flux of (5$\pm$2)$\times$10$^{-13}$\,{\ergs}, while
H$\alpha$ image data give 6.5 and 3.2$\times$10$^{-12}$\,{\ergs} 
\citep[][respectively]{bos02,ken79}. For this population, the unreddened 
EW(H$\alpha$) should be about 1200\AA. \citet{tor89} obtained an unreddened
value of 1175\AA, in good agreement with {\tt LavalSB} predictions.
The differences between the predicted and observed extinction, nebular
flux and stellar mass will be discussed in more details in \S\ref{2egen}.

\subsection{NGC\,5471}
\label{n5471}

NGC\,5471 is another GEHR in M\,101 more compact than NGC\,5461 and NGC\,5447
and may contain about 19 star clusters according to \citet{chen05}.
Most of the H$\alpha$ emission of this {\hii} region would
have been included within the LWRS aperture of FUSE. Unfortunately, this
{\hii} region has been observed using the MDRS aperture 
(4.0$^{\prime\prime}\times$20$^{\prime\prime}$), which implies that 
some OB stars are not included in the FUV spectrogram presented 
here (see Fig.~1). Also, in the FUSE data, no flux has been obtained 
in detector~2, which affects the quality of the synthesis since 
the LiF2A segment (which falls on the missing detector) is important
for the S/N of {\pv} and {\ciii} lines \citep[see][]{sah00}.
The FUSE spectrogram is shown in Figure~3c with S/N=9. The {\ciii} line 
profile displays no obvious wind feature. The best-fit model is obtained 
for a stellar population of 4.5$\pm$0.5\,Myr at 0.4\,{\zsun}. At 
0.1\,{\zsun}, a modeled stellar population of 3.5-4.0\,Myr can also 
reproduce the observed line profiles. Because of the noise, a standard 
IMF as been assumed. \citet{mas99} deduced an age 
of 2.9\,Myr for NGC\,5471, which is too young to explain the 
faint P\,Cygni profiles observed in the FUV diagnostic lines. Oxygen 
abundances ranging from 8.0 to 8.2 
\citep[0.2-0.3\zsun][]{tor89,ros94,mas99,bos02} are found in the 
literature, which is in good agreement with FUV synthesis. 

Adopting the 0.4\,{\zsun} best-fit model, the comparison between the 
observed and modeled continuum slopes indicate a low extinction, 
smaller than the uncertainties of 0.02. The FUV flux level suggests a 
stellar mass of (7$\pm$1)$\times$10$^{4}$\,{\msun} for NGC\,5471. 
\citet{mas99} obtained an extinction of 0.07 in the UV range, which is 
slightly higher than the FUV extinction. The FUV stellar mass deduced is 
consistent with the mass of 1.2$\times$10$^{5}$\,{\msun} from 
\citet{mas99}, considering the smaller aperture used with FUSE. 
Predictions reported in Table~3 are difficult to compare with the literature 
because of large differences 
between apertures. However, the FUV flux prediction is always below 
the values given from larger apertures \citep[e.g.][]{ken79,bos02}. 
\citet{tor89} measured a dereddened EW of 575\AA\ for H$\alpha$, consistent
with the predictions.

\subsection{NGC\,5458}
\label{n5458}

NGC\,5458 is an {\hii} region smaller and fainter than the previous ones 
in M\,101 and not much studied except for its X-ray source 
\citep{wan99,pen01,col04}. The FUSE spectrogram is presented in Figure~3d, 
and shows a S/N$\sim$10. The spectrogram displays photospheric profiles 
without evident signs of winds in both {\pv} and {\ciii} features. 
Sub-solar metallicity
models produce stellar line depths too weak compared to the observations.
The best-fit model is obtained for a 5.5-6.0\,Myr old stellar population
at {\zsun}. A standard IMF has been assumed since the line profiles are
less sensitive to the IMF when evolved O stars have disappeared. 
The continuum slope indicates a low extinction, below the uncertainties 
of 0.02. The flux level leads to a stellar mass of (1.1$\pm$0.4)$\times$10$^5$\,{\msun}. 
Other predicted observable parameters for NGC\,5458 are reported in Table~3.

\section{Discussion}

The massive stellar contents of several GEHR have been studied in detail
using the FUV spectral synthesis. The section below 
focuses on the global characteristics of the whole sample to better 
understand the physics of GEHR in general as well as the synthesis 
technique in the FUV.

\subsection{FUV Synthesis of Small Stellar Populations}
\label{mass}

Spectral synthesis is a powerful technique to obtain a good estimate
of the general characteristics of young integrated stellar populations. 
However, this technique usually assumes that the stars follow
an analytical IMF, and that the stars properly fill each bin of the mass 
function. But how high does the mass of the population must be in order to
be accurately described by an analytical IMF? The FUV is a good wavelength 
range to estimate this minimal mass for young systems. The FUV is 
especially sensitive to IMF statistical fluctuations at high masses 
since only O and B\,stars produce many photons below 1200\,\AA. 
Also, GEHR are very young systems and the disappearance of the most 
massive stars does not significantly affect the total stellar mass of 
the system.

From FUV synthesis of GEHR in M\,33 and M\,101 (\S\ref{syn}), it appears 
that a stellar mass greater than 1$\times$10$^3$\,{\msun} is needed to 
properly fulfill 
the IMF bins. As shown by NGC\,592, NGC\,604 (LWRS), and GEHR in M\,101,
a stellar mass of $\sim$1$\times$10$^4$\,{\msun} does not seem to suffer 
much of a statistical bias. However, the FUV synthesis of NGC\,604 
(MDRS), NGC\,595, and NGC\,588-NW reveals that a stellar mass closer to
$\sim$1$\times$10$^3$\,{\msun} becomes too low to obtain reliable
values of the age and mass of the star cluster because the
stellar line profiles are not those of a standard modeled population,
but those of a mix of a limited number of bright stars.
Note that the mass limit needs to be higher for younger systems, where
the dominant stars are of earlier spectral types than those found in a
slightly older population. This is because a younger population needs
to better fill the IMF higher mass bins and a more massive total stellar 
population is thus required. 

\citet{cer04} studied this problem from a theoretical point of view.
The lower mass limit of a few 10$^3$\,{\msun} found here for a 
synthesized population is fully consistent with their results, which 
suggest that the minimal initial cluster mass 
needed for synthesis modeling in the U-band is about 
8$\times$10$^3$\,{\msun} for a 5\,Myr population at 
0.4\,{\zsun}. Following their calculation, this minimal mass can 
be slightly lower at shorter wavelengths like the FUV range.
Using HST images where the stars of NGC\,588 were resolved, \citet{jam04}
obtained a standard IMF slope of 2.37$\pm$0.16 for NGC\,588 by using a
star counting technique and estimated a stellar mass of 
(5.8$\pm$0.5)$\times$10$^{3}$\,{\msun}, consistent with \citet{cer04} and
FUV synthesis. FUSE spectral synthesis of GEHR has clearly shown that their 
calculation not only applies to color bands, but also to stellar 
line profiles.

\subsection{The Flat IMF slope of GEHR}

The stellar IMF is a matter of debate since the work of \citet{sal55}. 
The generally 
accepted slope\footnote{A slope of 2.35 is traditionally called a 
Salpeter slope. However, this terminology is not appropriate for
stellar masses covered by FUSE since the work of \citet{sal55} 
applies to a lower mass range.}
for the massive OB star regime at all metallicities in every kind of 
environment (starbursts as well as star clusters),  is $\alpha$=2.35 
\citep[e.g.][]{mas98,sch00,gre04,pis04}.
However, the IMFs of GEHR derived from FUV line profiles seem to 
favor a relatively flat slope (see Table~2). Since FUV stellar flux is produced 
only by O and B stars, a small change in their relative numbers can affect 
the derived IMF slope. This result cannot be associated to
a bias due to the FUV synthesis since several, and bigger, young populations 
have been studied with the same technique and did not show such a 
flat slope \citep{pel04}. 

Some hypotheses could physically explain a flat IMF in the FUV range. 
One hypothesis is that B-type stars could still be more 
extinguished by dust than earlier type stars. If so, it would then
be more difficult to see them in the FUV, producing an artificially 
flatter IMF. However, the extinction values of individual stars in 
NGC\,604 obtained by \citet{bru03} do not show a significant correlation
with the spectral type, suggesting that B stars are not systematically
more extinguished than O stars. 

Another more plausible possibility is that 
the massive stars fill the IMF high mass bins relatively well, 
but not perfectly.  If some spectral types have slightly deviant
numbers from the analytical IMF, it will slightly change the 
integrated stellar line profiles in the same direction as NGC\,595 or 
NGC\,604-MDRS, i.e. by accentuating the integrated wind profiles.
Since a flatter IMF also produces more pronounced P\,Cygni profiles, it 
would be hard to differentiate the two cases. Consequently, even if the
population synthesis gives reliable and precise results on most physical 
parameters of the population (age, mass, metallicity, colors, fluxes) 
for a $>$1$\times$10$^3$\,{\msun} population (\S\ref{mass}), it appears 
that the stellar IMF slope derived from the FUV line profiles is a sign of a 
non-perfect filling of the IMF high mass bins.
This last possibility is supported by the IMF obtained from the star 
counting technique of \citet{jam04} on NGC\,588. They derived a standard 
IMF slope, but their IMF histogram clearly shows that some mass bins, 
especially at higher masses, are clearly deviant from the analytical slope.

\subsection{A second generation of stars in NGC\,5461}
\label{2egen}

The spectral synthesis of FUSE data on NGC\,5461 has predicted much lower
values for the H$\alpha$ flux (factor of 10), the stellar mass (factor of 
2 to 10), and the extinction than has been reported in the literature.
These discrepancies are hard to explain since most of the H$\alpha$ 
emission is included within the FUSE aperture. One plausible explanation 
is the presence of a second generation of stars in NGC\,5461, like the 
one observed in the LMC Cluster N11 \citep{wal92}. 
In the case of the star-forming region N11,
the central region is composed of a 3.5\,Myr stellar 
population which dominates the UV flux.  A surrounding nebulae is 
excited by a younger generation of stars which is not observed at 
short wavelengths because it is heavily reddened \citep{wal92}. 

The presence of a second generation of stars in NGC\,5461, younger 
and consequently more extinguished than the first one, could explain the larger 
extinction deduced at longer wavelengths, the stellar mass discrepancy as
well as the excess in nebular emission. The second generation cluster must 
then be relatively massive to explain the large differences in flux and mass.
It is not excluded that younger stars from different clusters are present rather
than a single second generation. 
Although there is no proof of such a population within NGC\,5461, 
this {\hii} region is a good candidate to host very massive stars,
younger than those actually detected with FUSE. 

It is also possible that younger stars are present within other GEHR
studied here. Unfortunately, because the FUSE aperture does not always 
include the whole system, it is impossible to confirm here if the 
difference between the predicted and observed H$\alpha$ fluxes comes 
from a second generation or not, as it is the case for NGC\,604 for 
example. Considering the detailed work of \citet{maiz04} on the 
attenuation maps of NGC\,604, the differences in H$\alpha$ fluxes 
and stellar masses in GEHR, including NGC\,5461, might also be due, 
at least partly, to the complexity of the gas and dust spatial distribution.

\section{Summary}

The evolutionary spectral synthesis technique in the FUV has been used to 
study the massive stellar content of nine GEHR in M\,33 and M\,101. 
Stellar masses, internal extinctions, and ages have been obtained 
for most of them. The comparison of the FUV synthesis results with values 
obtained from previous available works in various wavelength ranges has 
shown that the technique is reliable in most cases. 
The comparison of the GEHR with each other has confirmed
observationally that the synthesis technique must be applied to stellar 
populations of at least a few 10$^3$\,{\msun} in the FUV to avoid 
statistical fluctuations of the high mass end of the stellar IMF.
It has also revealed that a flat IMF slope is apparently favored for 
GEHR in the FUV, which is likely the first apparent effect of 
statistical fluctuations of the IMF for low mass populations.
FUV data suggests that giant {\hii} regions reach their maximum
nebular luminosity around 3.0-3.5\,Myr, coincident with the WR phase.
Finally, the {\hii} region NGC\,5461 in M101 is a good candidate to host
a second generation of stars more extinguished than, and formed after the 
cluster actually detected with FUSE.

\acknowledgments

The author warmly thanks N. R. Walborn and L. Drissen for very 
helpful comments that considerably improved the scientific content. 
This work was supported by NASA Long-Term Space Astrophysics grant NAG5-9173.

\begin{figure}
\begin{minipage}{5.4cm}
\includegraphics[trim=0mm 0mm 0mm 
0mm,clip,width=5.4cm]{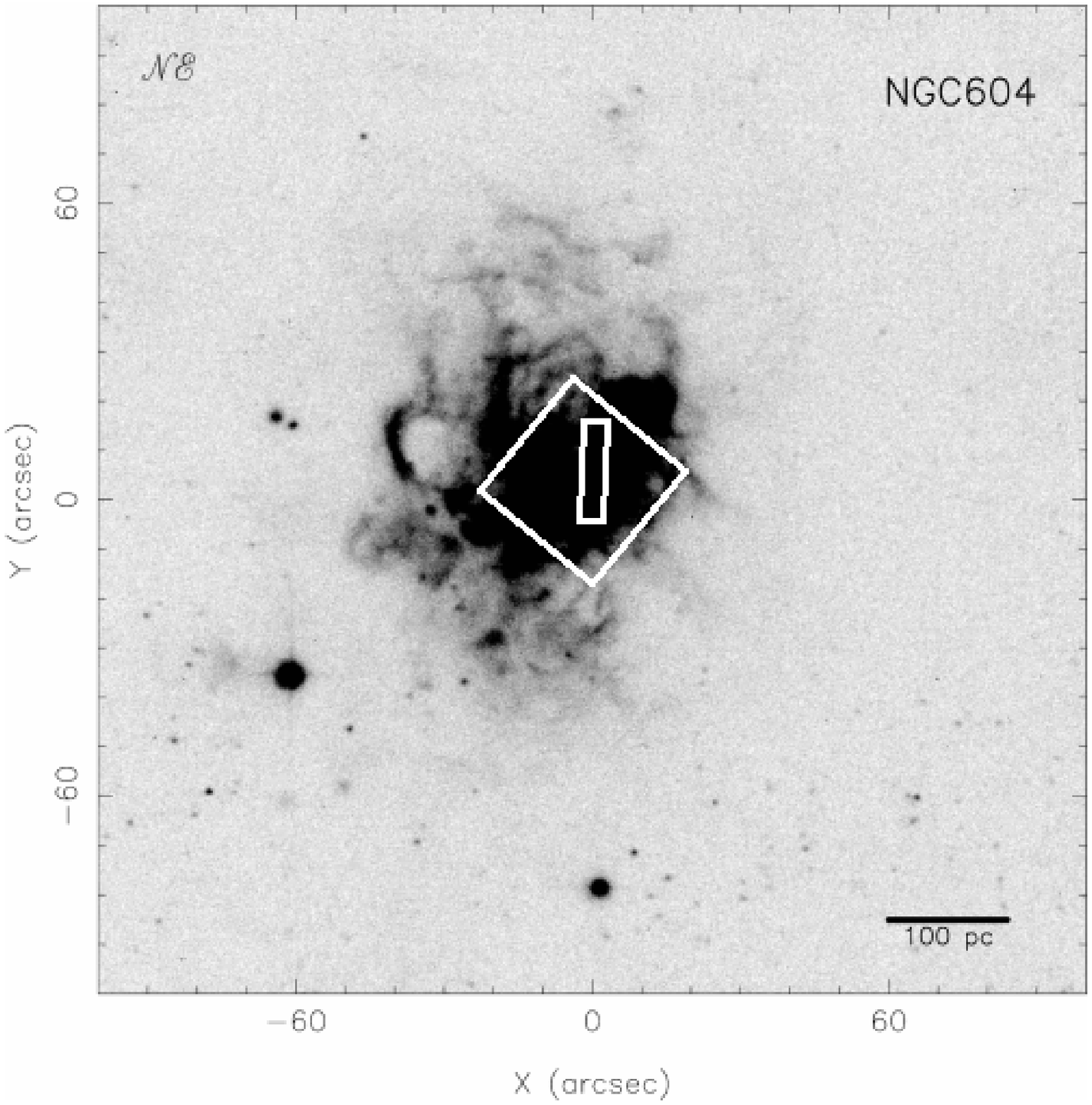} 
\includegraphics[trim=0mm 0mm 0mm 
0mm,clip,height=5.4cm]{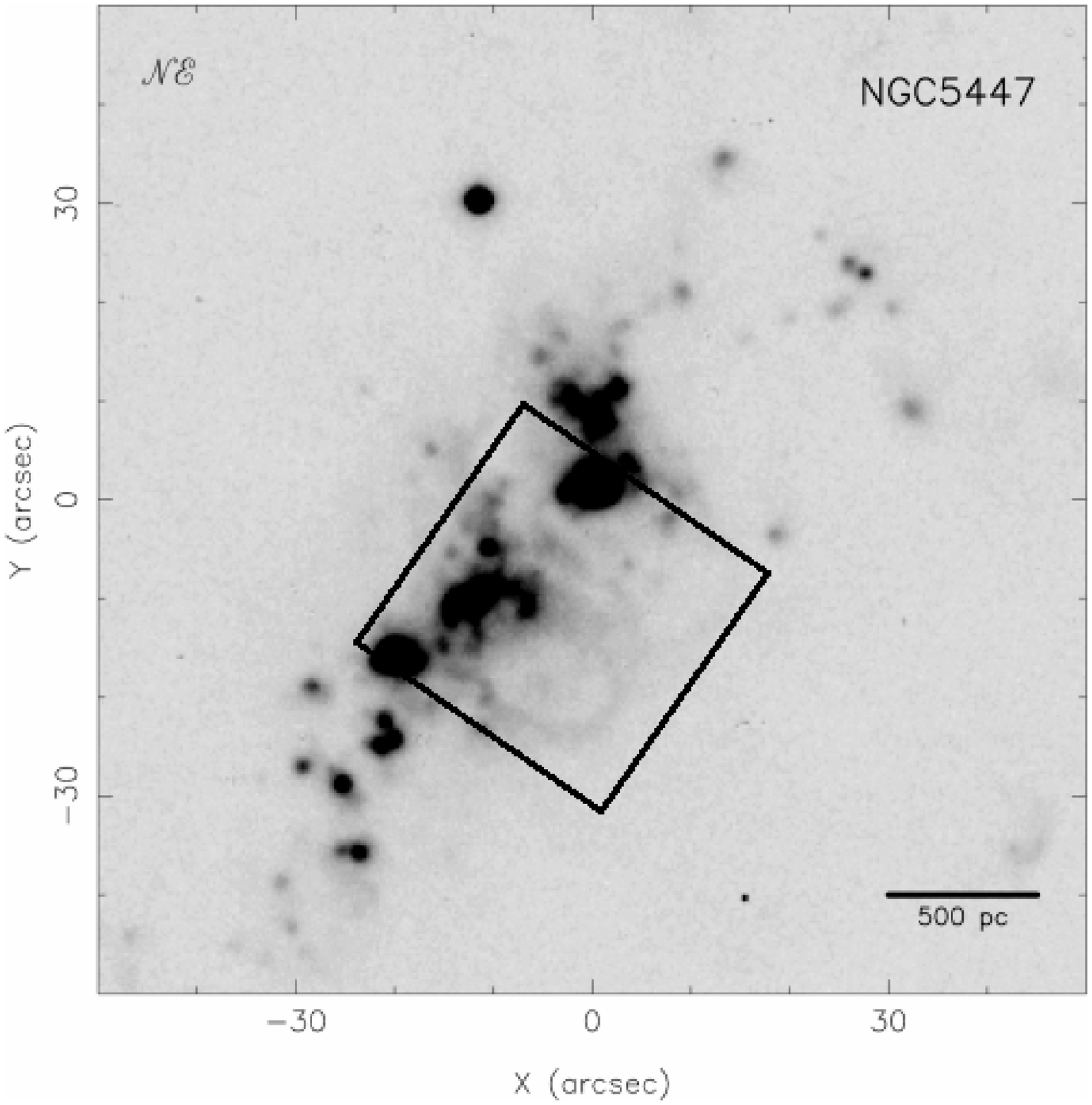} 
\includegraphics[trim=-20mm -20mm 0mm
0mm,clip,height=5.4cm]{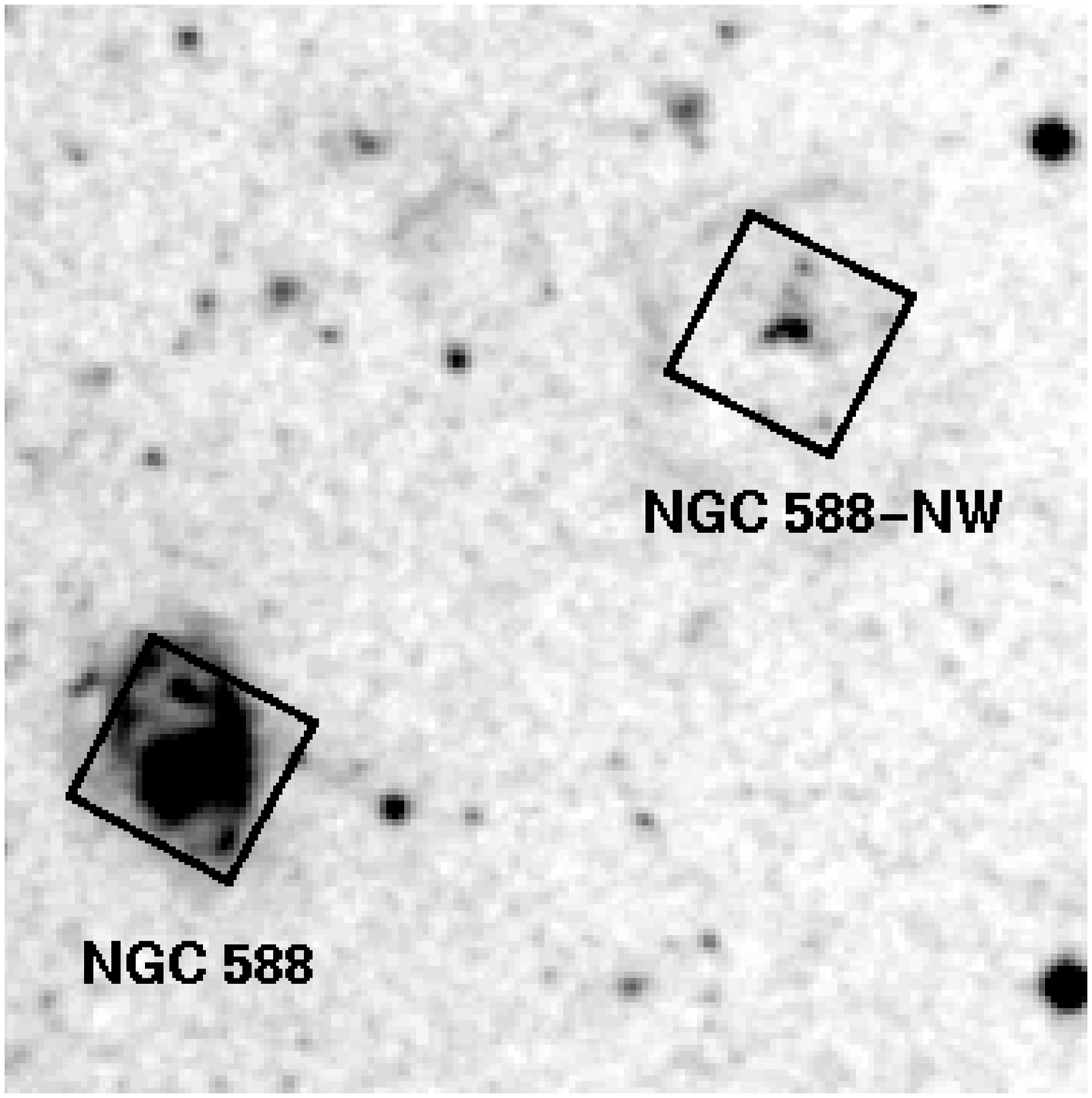} 
\end{minipage}
\begin{minipage}{5.4cm}
\includegraphics[trim=0mm 0mm 0mm 
0mm,clip,width=5.4cm]{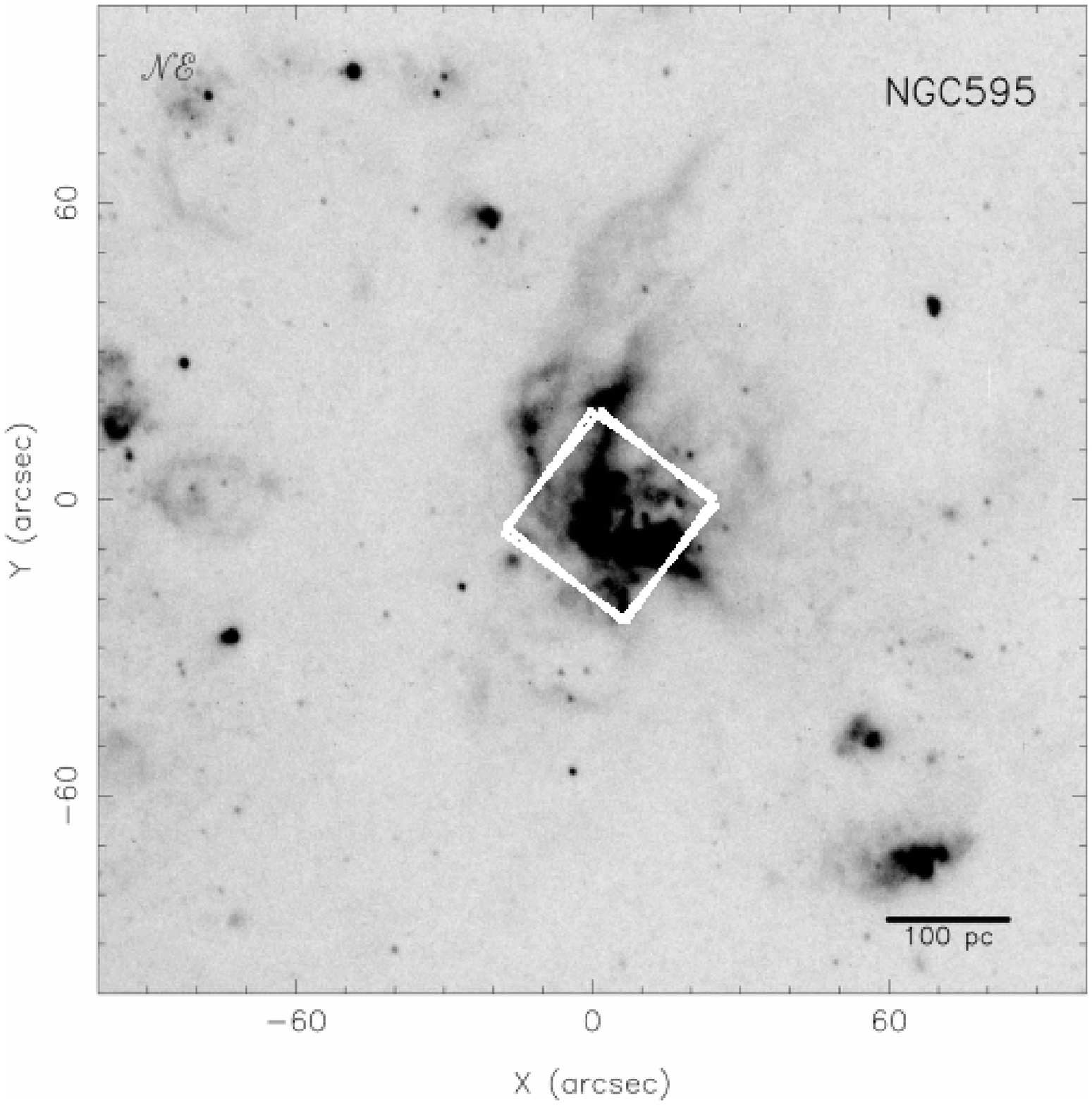} 
\includegraphics[trim=0mm 0mm 0mm 
0mm,clip,width=5.4cm]{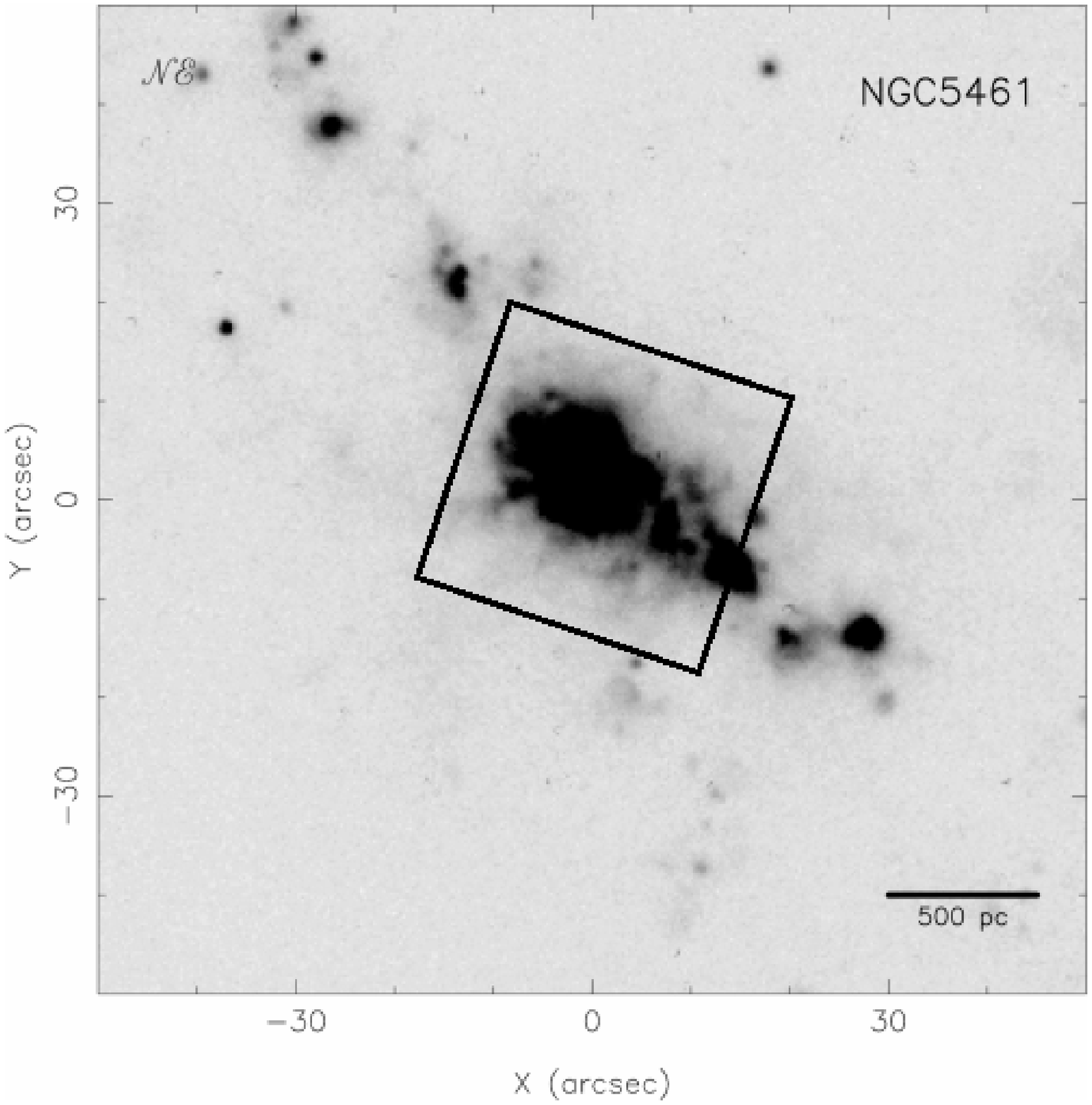} 
\includegraphics[trim=-20mm -20mm 0mm
0mm,clip,width=5.4cm]{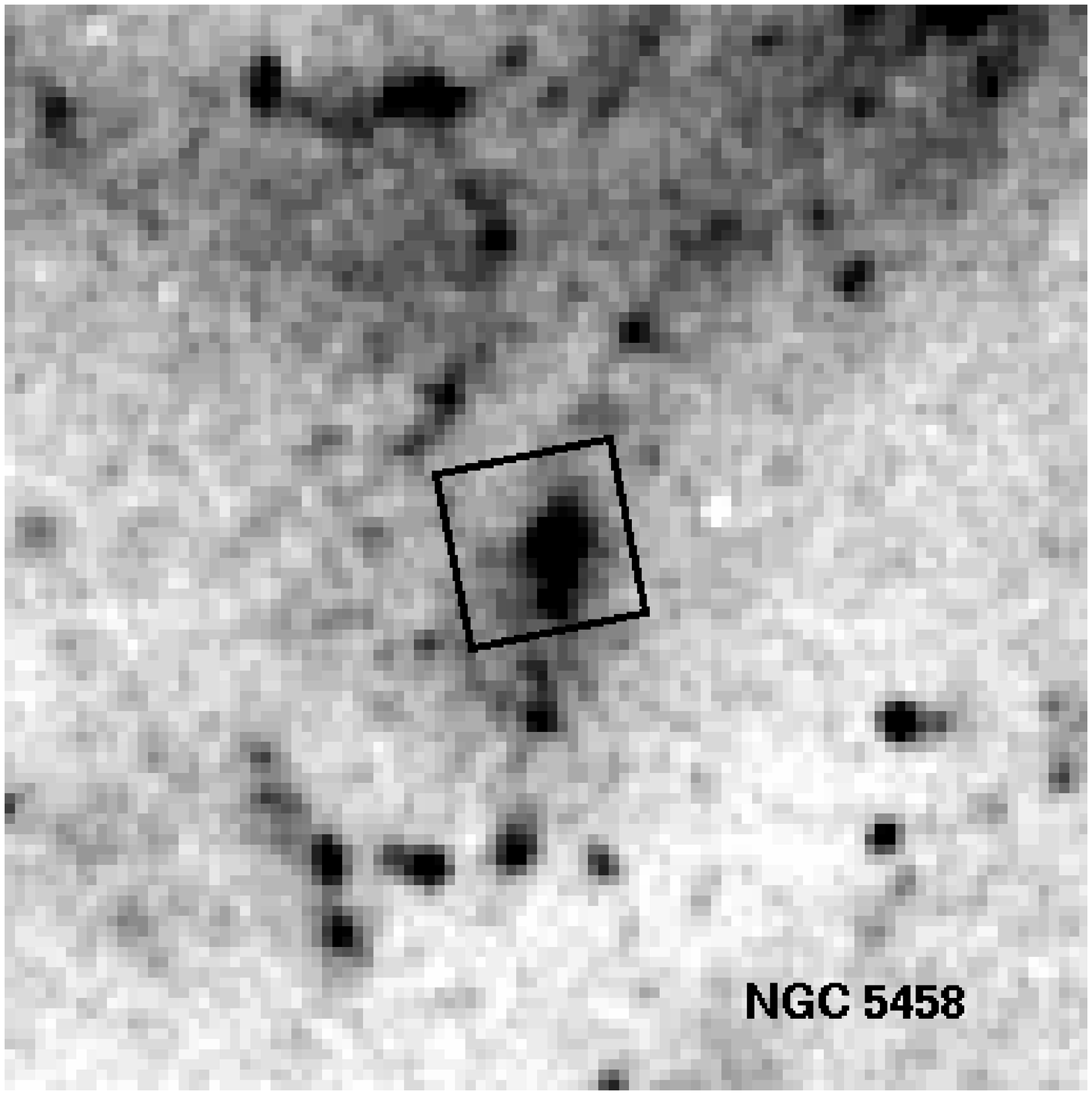} 
\end{minipage}
\begin{minipage}[t][-5.5cm][c]{5.4cm}
\includegraphics[trim=0mm 0mm 0mm 
0mm,clip,width=5.4cm]{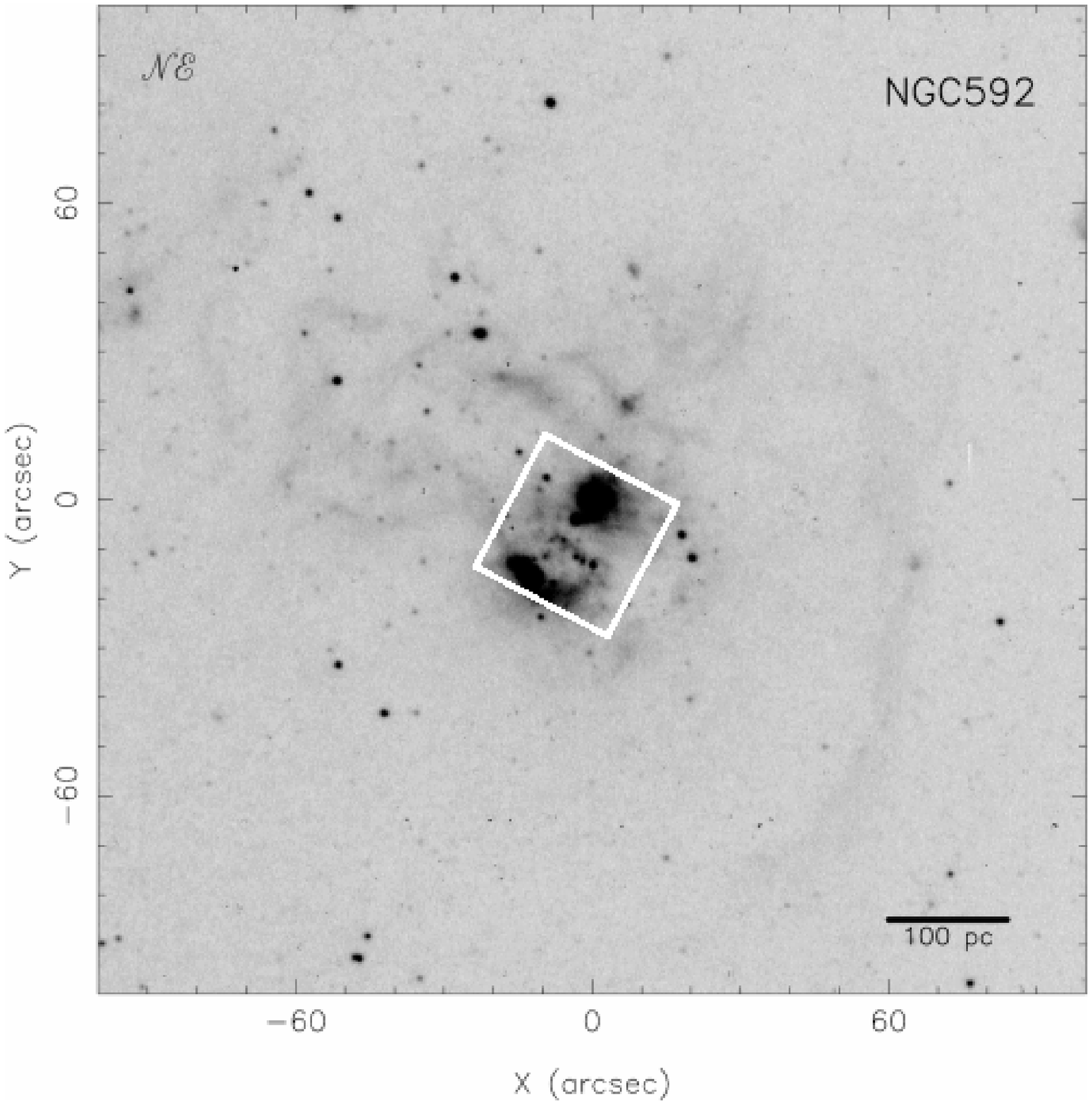} 
\includegraphics[trim=0mm 0mm 0mm 
0mm,clip,width=5.4cm]{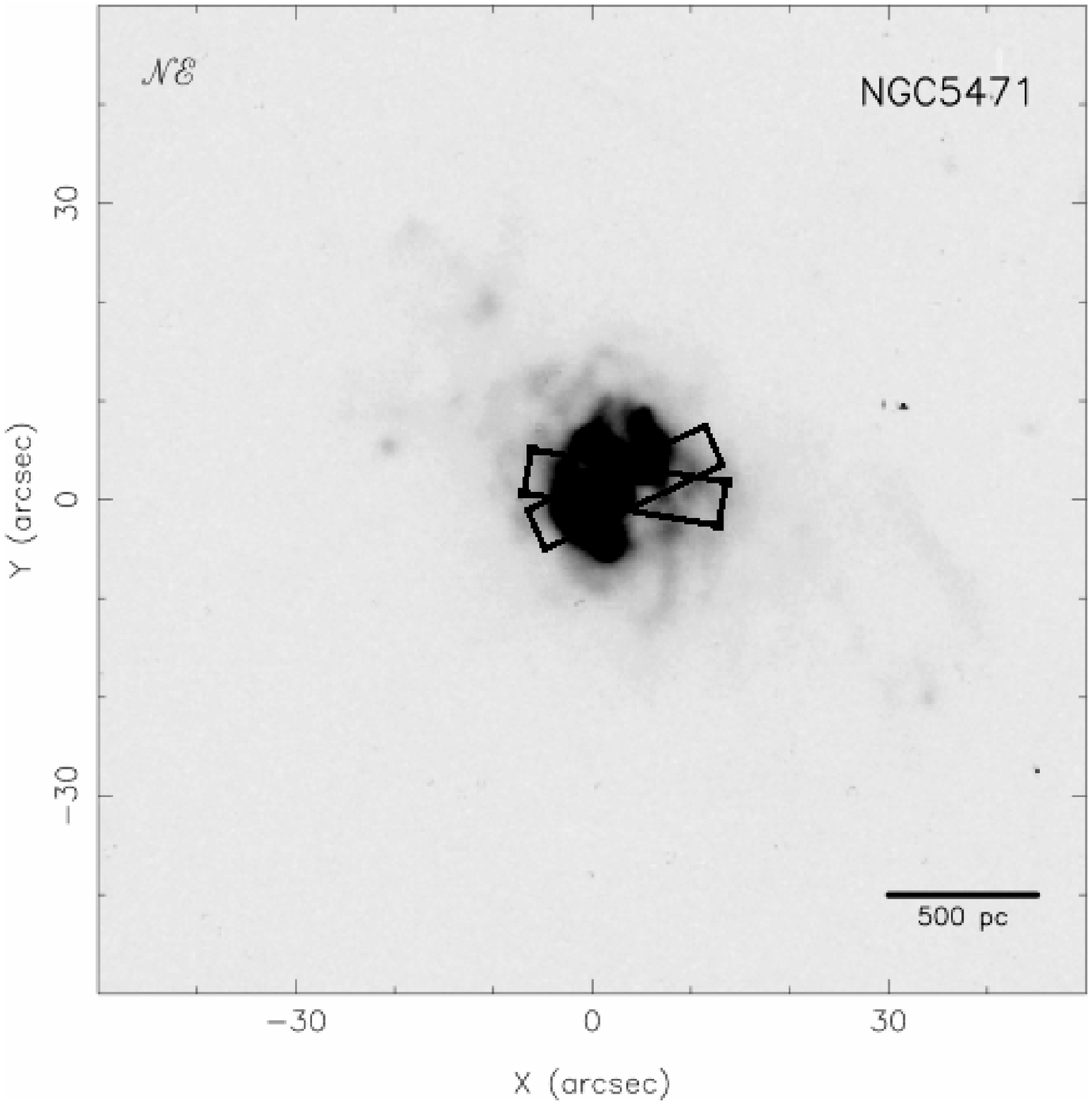} 
\end{minipage}
\caption{FUSE aperture locations. All images are from \citet{bos02} 
in H$\alpha$ (narrow-band filter) except for NGC\,588 and NGC\,5458 
from DSS. Squares represent the LWRS aperture 
(30$^{\prime\prime}\times$30$^{\prime\prime}$) and the rectangles 
represent the MDRS aperture (4$^{\prime\prime}\times$20$^{\prime\prime}$). 
North is up, east is left.}
\end{figure}

\begin{figure}
\plotone{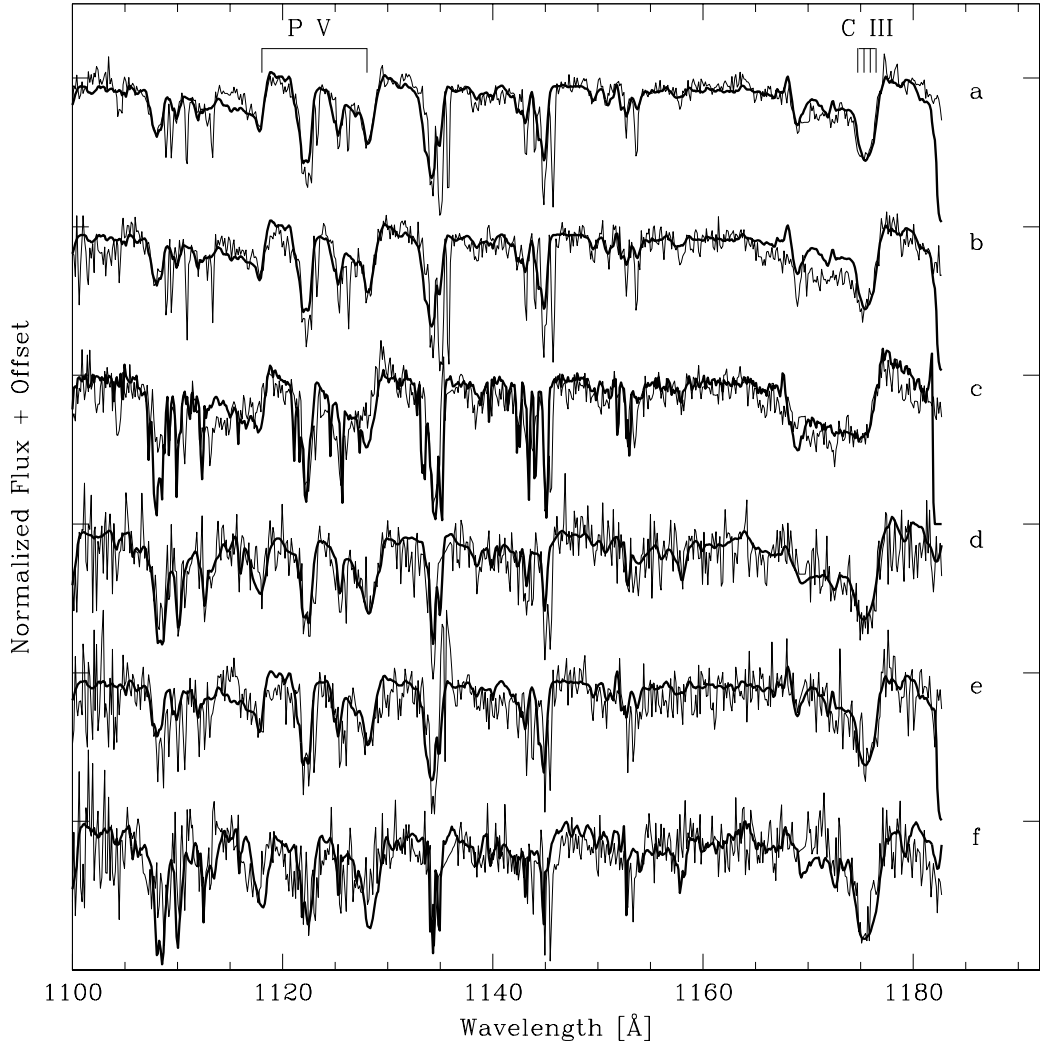}
\caption {FUV spectrograms of five GEHR in M\,33 superimposed on the {\tt LavalSB} best-fit model of a stellar population. a) NGC\,604 through the LWRS aperture and a 3.3\,Myr model at 0.4\,{\zsun} with $\alpha$(IMF)=1.5. b) NGC\,604 through the MDRS aperture and the same model as previously. c) NGC\,595 with the spectrogram of an O7\,I LMC star. d) NGC\,592 and a 4.0\,Myr model at {\zsun} with $\alpha$(IMF)=2.35. e) NGC\,588 and a 3.5\,Myr model at 0.4\,{\zsun} with $\alpha$(IMF)=1.5. f) NGC588-NW and the spectrogram of an O9.5\,III Galactic star. Thin lines: FUSE spectra. Thick lines: synthetic models from {\tt LavalSB} or stellar spectrograms. See text and Table~2 for details on the best-fit models.}
\end{figure}

\begin{figure}
\epsscale{.90}
\plotone{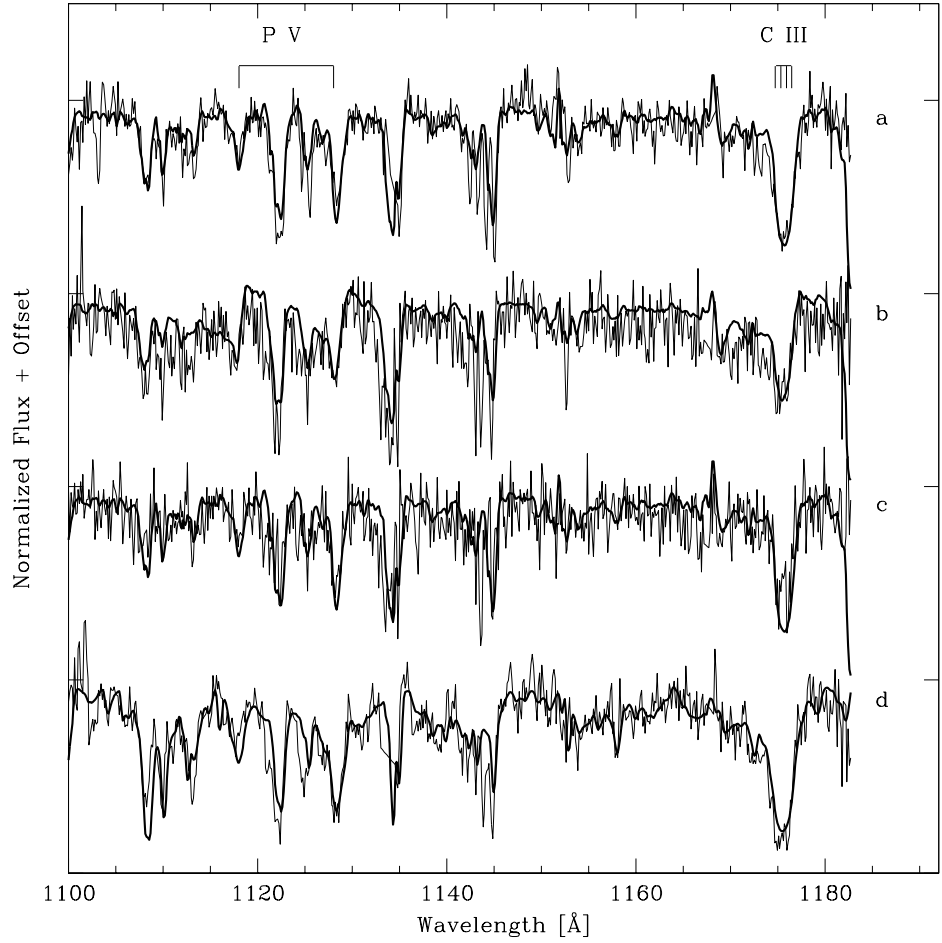}
\caption {FUV spectrograms of four GEHR in M\,101 superimposed on the {\tt LavalSB} best-fit model of a stellar population. a) NGC\,5447 and a 4.5\,Myr model at 0.4\,{\zsun} with $\alpha$=2.35. b) NGC\,5461 and a 3.3\,Myr model at 0.4\,{\zsun} with $\alpha$=1.5. c) NGC\,5471 and a 4.5\,Myr model at 0.4\,{\zsun} with $\alpha$=2.35. d) NGC\,5458 and a 5.5\,Myr model at {\zsun} with $\alpha$=2.35. Thin lines: FUSE spectra. Thick lines: synthetic models from {\tt LavalSB}. See text and Table~2 for details on the best-fit models.}
\end{figure}


\begin{deluxetable}{lccclrr}
\tabletypesize{\scriptsize}
\tablecaption{FUSE Observational Parameters of GEHR in M\,33 and M\,101}
\tablecolumns{6}
\tablehead{ \colhead{H{\sc {ii}} Region} & \colhead{$\alpha$(J2000)} & \colhead{$\delta$(J2000)} & \colhead{FUSE} & \colhead{Aperture\tablenotemark{a}} & \colhead{v$_{hel}$}& \colhead{F(1150)\tablenotemark{b}} \\
 & $^h$ $^m$ $^s$ & $^{\circ}$ $^{\prime}$ $^{\prime\prime}$ & \colhead{Program} &  & \colhead{[km s$^{-1}$]} & \colhead{[ergs s$^{-1}$ cm$^{-2}$ \AA$^{-1}$]} } 
\startdata
NGC\,588  & 01 32 45.5 & +30 38 55 & A08604 & LWRS & $-$174 & (10$\pm$1)$\times$10$^{-14}$ \\
NGC\,588-NW & 01 32 37.7 & +30 40 06 & A06105 & LWRS &        & (5.1$\pm$0.5)$\times$10$^{-14}$ \\
NGC\,592  & 01 33 12.3 & +30 38 49 & A08602 & LWRS & $-$162 & (10$\pm$1)$\times$10$^{-14}$ \\
NGC\,595  & 01 33 33.6 & +30 41 32 & A08603 & LWRS & $-$174 & (9.2$\pm$0.9)$\times$10$^{-14}$ \\
          & 01 33 33.6 & +30 41 32 & B01801 & LWRS &        & (10$\pm$1)$\times$10$^{-14}$ \\
NGC\,604  & 01 34 32.5 & +30 47 04 & A08601 & LWRS & $-$226 & (100$\pm$10)$\times$10$^{-14}$ \\
          & 01 34 32.4 & +30 47 04 & B01802 & MDRS &        & (16$\pm$2)$\times$10$^{-14}$ \\
NGC\,5447 & 14 02 28.6 & +54 16 11 & B01803 & LWRS &  152 & (8.4$\pm$0.8)$\times$10$^{-14}$ \\
NGC\,5458 & 14 03 12.6 & +54 17 55 & C07001 & LWRS &  241 & (3.8$\pm$0.5)$\times$10$^{-14}$ \\
NGC\,5461 & 14 03 41.3 & +54 19 05 & A08605 & LWRS &  298 & (6.3$\pm$0.7)$\times$10$^{-14}$ \\
NGC\,5471 & 14 04 28.7 & +54 23 49 & B01805 & MDRS &  297 & (4.2$\pm$0.6)$\times$10$^{-14}$ 
\enddata

\tablenotetext{a}{ The LWRS aperture is 30$^{\prime\prime}\times$30$^{\prime\prime}$ and MDRS aperture is 4$^{\prime\prime}\times$20$^{\prime\prime}$.}
\tablenotetext{b}{ Not corrected for extinction.}

\end{deluxetable}


\begin{deluxetable}{lccrcr}
\tabletypesize{\scriptsize}
\tablecaption{FUV Synthesis Parameters of GEHR in M\,33 and M\,101}
\tablecolumns{6}
\tablehead{ \colhead{H{\sc {ii}} Region} & \colhead{Age} & \colhead{Z$_{syn}$} & \colhead{$\alpha$(IMF)} & \colhead{E(B-V)$_i$} & \colhead{Stellar Mass}  \\
 & \colhead{[Myr]} &  &  & $\pm$0.02 & \colhead{[\msun]} } 
\startdata
\multicolumn{6}{l}{\bf Best-Fit Parameters} \\
NGC\,588     & 3.5$\pm$0.5 & 0.4 & $<$2.35 & 0.06 & (1.3$\pm$0.6)$\times$10$^3$ \\
NGC\,588-NW & 5-6         & 0.4 &  -   & $\sim$0       &  $\sim$1$\times$10$^3$ \\
NGC\,592     & 4.0$\pm$0.5 & 1   & 2.35 & 0.07 & (1.0$\pm$0.3)$\times$10$^4$  \\
NGC\,595     & 3.5$\pm$0.5 & 0.4 &  -   & 0.02 & $\sim$1$\times$10$^3$  \\
NGC\,604-LWRS& 3.3$\pm$0.1 & 0.4 & 1.50 & 0.03 & (7$\pm$2)$\times$10$^3$   \\
NGC\,604-MDRS& 3.3$\pm$0.2 & 0.4 & 1.50 & 0.03 & (1.0$\pm$0.3)$\times$10$^3$  \\
NGC\,5447    & 4.5$\pm$0.5 & 0.4 & $\leq$2.35 & $\sim$0 & (1.2$\pm$0.2)$\times$10$^5$  \\
NGC\,5458    & 5.5$\pm$0.5 & 1   & 2.35 & $\sim$0 & (1.1$\pm$0.4)$\times$10$^5$   \\
NGC\,5461    & 3.3$\pm$0.2 & 0.4 & 1.50 & $\sim$0 & (1.5$\pm$0.4)$\times$10$^4$   \\
NGC\,5471    & 4.5$\pm$0.5 & 0.4 & 2.35 & $\sim$0 & (7$\pm$1)$\times$10$^4$ \\
\hline \\
\multicolumn{6}{l}{\bf Other Good-Fit Parameters} \\
NGC\,588     & 4.5$\pm$1.0 & 0.1 & $<$2.35 & \nodata & \nodata \\
NGC\,604-LWRS& 3.9$\pm$0.1 & 0.1 & 1.50 & 0.03 & (1.1$\pm$0.2)$\times$10$^4$  \\
NGC\,604-LWRS& 3.5$\pm$0.3 & 0.1 & 2.35 & 0.03 & (2.5$\pm$0.3)$\times$10$^4$  \\
NGC\,604-LWRS& 3.0$\pm$0.3 & 0.4 & 2.35 & 0.03 & (2.2$\pm$0.5)$\times$10$^4$  \\
NGC\,5461    & 4.0$\pm$0.5 & 0.1 & 1.50 & $\sim$0 & (2.2$\pm$0.6)$\times$10$^4$ \\
NGC\,5471    & 3.5-4.0     & 0.1 & 2.35 & $\sim$0 & \nodata 
\enddata

\end{deluxetable}


\begin{deluxetable}{lrrrrrc}
\tabletypesize{\scriptsize}
\tablecaption{Predicted Parameters for GEHR from FUV Synthesis}
\tablecolumns{8}
\tablehead{ \colhead{H{\sc {ii}} Region} & \colhead{\#O} & \colhead{\#WR} & \colhead{F(H$\alpha$)\tablenotemark{a}} & \colhead{F(5500)\tablenotemark{a}} & \colhead{F(H$\alpha$)\tablenotemark{a}} & \colhead{Ref} \\
  &  &  & \colhead{Predicted} &  & \colhead{Observed} &  }
\startdata
NGC\,588     & 15 & 2 & 2.8$\times$10$^{-12}$ & 4$\times$10$^{-15}$ & 2-3$\times$10$^{-12}$     & 1, 2 \\
NGC\,588-NW  &  4 & 0 & 2$\times$10$^{-13}$   & 1$\times$10$^{-15}$ & \nodata                 & \nodata \\
NGC\,592     & 40 & 4 & 2.7$\times$10$^{-12}$ & 1$\times$10$^{-14}$ & 2.7$\times$10$^{-12}$     & 2 \\
NGC\,595     & 10 & 1 & 1.3$\times$10$^{-12}$ & 2$\times$10$^{-15}$ & 8.8-11$\times$10$^{-12}$  & 2, 1 \\
NGC\,604-LWRS& 90 & 9 & 2$\times$10$^{-11}$   & 3$\times$10$^{-12}$ & 3.3-4.0$\times$10$^{-11}$ & 3, 2\\
NGC\,604-MDRS& 12 & 2 & 2$\times$10$^{-12}$   & 3$\times$10$^{-13}$ & \nodata                 & \nodata \\
NGC\,5447    &460 &34 & 5.7$\times$10$^{-13}$ & 9.2$\times$10$^{-15}$ & 1.6-4.7$\times$10$^{-12}$ & 1, 2 \\
NGC\,5458    &150 &46 & 1.7$\times$10$^{-13}$ & 6$\times$10$^{-16}$ & \nodata                 & \nodata \\
NGC\,5461    &175 &18 & 5$\times$10$^{-13}$   & 7$\times$10$^{-16}$ & 3.2-6.5$\times$10$^{-12}$ & 1, 2 \\
NGC\,5471    &290 &20 & 3.5$\times$10$^{-13}$ & 6$\times$10$^{-16}$ & 3.5-6.2$\times$10$^{-12}$ & 1, 2
\enddata

\tablerefs{1. \citet{ken79}; 2. \citet{bos02}; 3. \citet{gonz00}.}
\tablenotetext{a}{Unreddened fluxes in \ergs.}

\end{deluxetable}

\end{document}